\documentclass[12pt,epsf]{article}

\usepackage{graphicx,float}
\usepackage{mathrsfs,array,multirow}
\usepackage{amssymb,amsfonts}
\usepackage{amsmath}
\usepackage{color}

\newcommand{\be}{\begin{equation}}

\newcommand{\ee}{\end{equation}}

\newcommand{\ba}{\begin{array}}

\newcommand{\ea}{\end{array}}

\begin{document}
\begin{titlepage}
\vspace{.5in}
\begin{flushright}
YITP-14-69
\end{flushright}
\vspace{0.5cm}

\begin{center}
{\Large\bf Oscillating Fubini instantons in curved space }\\
\vspace{.4in}

  {$\rm{Bum-Hoon \,\, Lee}^{\P\dag}$}\footnote{\it email:bhl@sogang.ac.kr}\,\,
  {$\rm{Wonwoo \,\, Lee}^{\P}$}\footnote{\it email:warrior@sogang.ac.kr}\,\,
  {$\rm{Daeho \,\, Ro}^{\dag}$}\footnote{\it email:dhro@sogang.ac.kr}\,\,
  {$\rm{Dong-han \,\, Yeom}^{\S\ddag}$}\footnote{\it email:innocent.yeom@gmail.com} \\

  {\small \P \it Center for Quantum Spacetime, Sogang University, Seoul 121-742, Korea}\\
  {\small \dag \it Department of Physics, Sogang University, Seoul 121-742,
  Korea}\\
  {\small \S \it Yukawa Institute for Theoretical Physics, Kyoto University, Kyoto 606-8502, Japan}\\
  {\small \ddag \it Leung Center for Cosmology and Particle Astrophysics,\\
   National Taiwan University, Taipei 10617, Taiwan}\\

\vspace{.5in}
\today
\end{center}

\begin{center}
{\large\bf Abstract}
\end{center}
\begin{center}
\begin{minipage}{4.75in}

{\small \,\,\,\, A Fubini instanton is a bounce solution which describes the decay of a vacuum state located at the top of the tachyonic potential {\it via} the tunneling without a barrier. We investigate various types of Fubini instantons of a self-gravitating scalar field under a tachyonic quartic potential. With gravity taken into account, we show there exist various types of unexpected solutions including oscillating bounce solutions. We present numerically oscillating Fubini bounce solutions in anti-de Sitter and de Sitter spaces. We construct the parametric phase diagrams of the solutions, which is the extension of our previous work. Of particular significance is that there always exist solutions in all parameter spaces in anti-de Sitter space. The regions are divided depending on the number of oscillations. On the other hand, de Sitter space allows solutions with codimension-one in parameter spaces. We numerically evaluate semiclassical exponents which give the finite tunneling probabilities.}


\end{minipage}
\end{center}
\end{titlepage}

%\newpage

%\tableofcontents

\newpage

\section{Introduction \label{Sec.1}}

A bounce solution represents an unstable nontopological configuration that corresponds to the saddle point configuration of the Euclidean action. The second derivative of the Euclidean action around the bounce has one negative eigenvalue which is related to the imaginary part of the energy. The bounce solution describes the decay of a metastable vacuum state and determines the semiclassical exponent of the tunneling probability \cite{col002, calcol}. One can use this probability to calculate the lifetime of the metastable vacuum state \cite{klph, kl00}.

There are three different kinds of bounce solutions, which have become a remarkable Euclidean solution as applied to cosmology \cite{guth, sato, adlin6, alste, hawking, lllo}. One corresponds to a vacuum bubble. The mechanism was introduced to describe a phase transition \cite{col002, voloshin} without gravity. The formalism was later developed with gravity \cite{bnu02} and extended to the case with an arbitrary vacuum energy \cite{par02}. The possible types of the true vacuum bubbles in de Sitter (dS) and general background were studied in Refs.\ \cite{llww, bcwc}, in which six types of the true vacuum bubble were analyzed in more detail. For the nucleation of a false vacuum bubble in the true vacuum background, the nucleation of a large false vacuum bubble in dS space was originally obtained in Ref.\ \cite{kimwein}. The possible types of the false vacuum bubbles were investigated. The false vacuum solutions only with compact geometry are possible in Einstein gravity \cite{bcwc}. The bounce solutions mediating tunneling between the degenerate vacua was also studied \cite{hawe, hhlc}.

Another corresponds to oscillating bounce solutions with oscillations around the minimum of the inverted potential. The crossing number of the potential barrier by an oscillating solution is denoted as $i$. With this convention an ordinary bounce solution corresponds to $i$ equal to $1$. The existence of oscillating solutions, $i>1$, is highly probable if the oscillating solutions are allowed and the contribution of those could be added. Little investigation has been carried out on the physical meaning of the oscillating solutions in Lorentzian spacetime \cite{lllo}. The study on the existence of the Euclidean solution deserves to receive attention in this stage. Let's see the tunneling problem in quantum mechanics. The action of the particle with $i$ times oscillation around the minimum of the inverted potential should be $i$ times of the ordinary bounce solution. Hence the $i=1$ contribution dominates the path integral \cite{zinn}. In four dimension, there is a damping term in the scalar field equation in the absence of gravity. Therefore we cannot expect the existence of oscillating solutions with $O(4)$ symmetry. There can only exist an ordinary bounce solution without oscillation. If the gravity is taken into account, the situation changes drastically. For instance, the role of a damping term in the scalar field equation can be changed to an antidamping term if the dS region is included during the transition. The oscillating solution in dS background was first studied in Ref.\ \cite{hawe}, in which the authors numerically found the oscillating bounces. The oscillating solutions in the symmetric double potential were obtained in the general background space, in which the condition for the existence of the oscillating solution was analyzed \cite{lllo}. The oscillating instantons as homogeneous tunneling channels was also studied \cite{bwd0}. The properties of the oscillating instantons were intensively studied in Ref.\ \cite{NeMode}.

The other corresponds to the so-called Hawking-Moss (HM) instanton \cite{hawking}. The solution describes the scalar field jumping simultaneously onto the top of the potential barrier. Thus, the tunneling occurs everywhere at the same time. Historically, the model with HM instanton was based on the phase transition under the Coleman-Weinberg mechanism of
symmetry breaking \cite{colwein}. Among three kinds of bounce solutions, the probabilities of oscillating instantons are smaller than those of Coleman-de Luccia instantons and hence it was the reason why oscillating instantons are overlooked in Einstein gravity. However, if we add correction terms to the gravity sector, then this may change the status of oscillating instantons. One interesting example is nonlinear massive gravity \cite{zss, syz, zsys}. In this case, the correction term from the gravity sector may enhance HM instantons (and hence probably oscillating instantons, too) though we need further investigation.

The Fubini instanton \cite{fubi01, lipatov, linde00} describes the decay of a metastable vacuum state by tunneling instead of a rolling down on the tachyonic potential consisted of a quartic term only. In a scale invariant Lagrangian field theory, the instanton introduces a fundamental scale of hadron phenomena by means of a dilatation noninvariant vacuum state. The conformal invariance of the scalar field theory allows the existence of the Euclidean solution with arbitrary size and the same probability. The explicit form of the solution was obtained in \cite{fubi01}. If the scale invariance breaks down due to the existence of the mass term a bounce solution does not exist. In other words, the particle cannot have enough energy to climb the hill up to $\Phi=0$ \cite{affle}. The Fubini instanton is a one-parameter family of bounce solutions representing tunneling without a barrier, which is interpolating between the state at the top of the potential and an arbitrary state. The solutions could be considered as a ball composed of only a thick wall except for one point at the center of the solution with an arbitrary state lower than the outer vacuum state, unlike a vacuum bubble that composed of an inside part with a lower vacuum state and a wall. When the gravity is taken into account, the conformal invariance is broken. However, the instanton solution was studied in a conformally invariant model in a fixed background without the backreaction \cite{gmst, khleb, loran01}. The tunneling without a barrier was studied in the flat potentials \cite{kmwe, kml, jest}. The solution representing the tunneling from the local maximum of the symmetric double well potential to one of minima of the potential was obtained in anti-de Sitter (AdS) space \cite{lllo}. Recently, the vacuum decay from the flat Minkowski to AdS space was studied as a tunneling without a barrier \cite{kss0}.

We have shown the existence of numerical solutions of the Fubini instanton in the initial flat and AdS spaces for the potential with only the quartic term \cite{bwdd}. The oscillating solutions under a double-hump potential have been studied in Ref.\ \cite{doublehump}. We have also shown numerically there exist solutions for the potential with both a quartic and a quadratic term irrespective of the value of the cosmological constant. We obtained the solutions with $Z_2$ symmetry in the dS background \cite{bwdd}. Recently, the oscillating bounce solutions under flat potential barriers was extensively studied in dS space \cite{balale}, in which the authors analyzed the variety of solutions using an instanton diagram \cite{bofl}. In the present paper, we investigate oscillating Fubini instantons in AdS and dS spaces constructing the parametric phase diagram.

In the model of cosmology, the first picture of the inflationary multiverse scenario was proposed to make the universe scenario without the cosmological singularity problem. The picture has the interesting property of self-reproducing or regenerating an exponentially expanding universe. In this scenario, the universe as a whole consists of different parts of inflationary domains or an infinite number of miniuniverses (bubbles) \cite{alinde00}. A very large class of inflationary scenarios have been analyzed leading to a regime called eternal inflation, in which once the inflation can start, it never stops globally. The scenario has the regions separated by more than an observable universe or a Hubble volume without correlation \cite{vil01, linde03, gu90}. Two scenarios could be combined to the eternally inflating multiverse scenario.

The cosmic landscape of string theory is the design that involves a huge number of different metastable and stable vacua, in which the vacua could be approximated by a set of fields and a potential. The space of all string theory vacua or these fields is called the landscape \cite{land00, land01, kklt}. In the landscape, the vacua could have a chance to obtain the appropriate value equal to that in our universe if the scenario could be realized. On the other hand, a supergravity from M-theory could have a dS maximum, which is unbounded from below \cite{klph, kl00, chull, klps}. If there exist various states corresponding to metastable and stable vacua, the tunneling could be interesting phenomena. To simply things, we could assume the potential has a lot of vacuum states. One of them could have a very high hill. Then the vicinity of the top of the hill could be approximated as a tachyonic potential. In the new inflationary model the potential could be also approximated to a tachyonic quartic potential when the scalar field is small value \cite{adlin6, alste, colwein}.

The instanton solutions have renewed interest in the AdS/CFT correspondence \cite{hapet, hapet00, bara}. The Fubini instanton under a tachyonic potential in AdS bulk could be related to an instanton solution under a tachyonic potential in the boundary conformal field theory. The ambiguity is what kind of instanton in the boundary corresponds to the bulk Fubini instanton.

The tunneling process is quantum phenomenon where a particle can penetrate through a finite potential barrier. The simplest case in quantum tunneling is a one-dimensional problem, which is extensively studied. However, the extension of the problem to a higher dimension is not straightforward. If the gravity is taken into account, the situation is much more complicated.

In these perspectives, the tunneling phenomenon including the effect of gravity is worthwhile to be studied in more detail. The purpose of this paper is to investigate further this tunneling process by finding the diversity of tunneling solutions and construct the parametric phase diagrams of oscillating bounce solutions.

The outline of this paper is as follows: In the next section we set up the basic framework for this paper. We explain the boundary conditions for the numerical solutions. We employ the potential with only the quartic self-interaction term. In Sec.\ \ref{Sec.3}, we present numerical solutions including oscillating bounce solutions. For the decay probability, we evaluate the action difference between the action of bounce solution and that of the background by numerical calculation. In Sec.\ \ref{Sec.4}, we construct the parametric phase diagram in AdS and dS spaces, which is the extension of our previous work \cite{bwdd}. In the parametric phase diagrams, the solutions occupy the area composed of two parameters in AdS space, while the solutions occupy the line in dS space. In the final section, we summarize and discuss our results. In the Appendix, we numerically prove the finiteness of the exponent $B$ to give the finite probability.

\section{Setup \label{Sec.2}}

The law of exponential decay is a good approximation to describe quantum tunneling phenomena. In the semiclassical approximation, the decay probability coming from the imaginary part of the energy of a metastable vacuum state is represented as $Ae^{-B}$. The prefactor $A$ is a functional determinant evaluated from the Gaussian integral over fluctuations around the classical solution \cite{calcol, ccol0, ccol1, ccol2}. The exponent $B=S^{bs}_E-S^{bg}_E$ is the difference between the Euclidean action of the bounce solution and the background action. We are interested in finding the exponent $B$.

We consider the tunneling phenomena in Einstein gravity minimally coupled to a scalar field with a tachyonic potential. We consider the action
\begin{equation}
S=  \int_{\mathcal M} \sqrt{-g} d^4 x \left[ \frac{R}{2\kappa}
-\frac{1}{2}{\nabla_\mu}\Phi {\nabla^\mu}\Phi -U(\Phi)\right]
+ \oint_{\partial \mathcal M} \sqrt{h} d^3 x \frac{K-K_o}{\kappa}\,,
\label{f-action}
\end{equation}
where $g=\det g_{\mu\nu}$, $\kappa \equiv 8\pi G$, $R$ denotes the
scalar curvature of the spacetime $\mathcal M$, and $h$ is the determinant of the first fundamental form. $K$ and $K_o$ are the traces of the second fundamental form of the boundary $\partial \mathcal M$ for the metric $g_{\mu\nu}$ and $\eta_{\mu\nu}$, respectively. The second term on the right-hand side is the so-called York-Gibbons-Hawking boundary term \cite{York, giha}. Here we adopt the sign conventions in Ref.\ \cite{misner}.

We consider the tachyonic potential with only a quartic self-interaction term as in Ref.\ \cite{bwdd}
\begin{equation}
U(\Phi)= -\frac{\lambda}{4}\Phi^4 + U_o,
\end{equation}
where the coupling constant $\lambda > 0$. $U_o$ is related to the cosmological constant as $\Lambda=\kappa U_o$. The background space will be dS, flat and AdS depending on the values of $U_o$. This potential is unbounded from below on either side of the center.

We assume an initial field configuration on the top of the potential, in which the field expectation value is spatially homogeneous and equal to zero. This configuration on top of the tachyonic quartic potential can be a metastable vacuum state \cite{fubi01}. The field has a finite probability to leave the top to an arbitrary state inhomogeneously by tunneling instead of a rolling down on the tachyonic potential. In what follows, we explore the transition process through the nucleation of Fubini instanton. We employ the Euclidean path integral approach for the transition probability. The semiclassical approximation leads to the classical equation of motion of a single particle.

We assume Euclidean $O(4)$-symmetry for the dominant contribution to the decay probability. The geometry is then written as
\begin{equation}
ds^{2}\ =\ d\eta^{2}+\rho(\eta)^2\left[d\chi^{2}+\sin^{2}\chi\left(d\theta^{2}+\sin^{2}\theta d\phi^{2}\right)\right]. \label{o4 symmetry}
\end{equation}
The scalar field $\Phi$ as well as $\rho$ depends only on $\eta$. The field equations turn out to be
\begin{equation}
\Phi''+\frac{3\rho'}{\rho}\Phi' =- \frac{d(-U)}{d\Phi}, ~~~
\rho'' = -\frac{\kappa}{3}\rho\left({\Phi'}^{2}+U\right)\,, \label{eom2}
\end{equation}
and the Hamiltonian constraint is given by
\begin{equation}
{\rho'}^{2} - 1 - \frac{\kappa\rho^{2}}{3} \left(\frac{1}{2}{\Phi'}^{2}-U\right) =0, \label{eom3}
\end{equation}
where the prime denotes differentiation with respect to $\eta$. The first equation in Eq.\ (\ref{eom2}) is formally equal to a one-particle equation of motion in the inverted potential in Newtonian mechanics. The second term on the left-hand side can be interpreted as a damping term. It can play the role of an antidamping term if $\rho'$ is negative as in dS space.

To solve the equations of motion, we should impose appropriate boundary conditions. In the absence of gravity, the boundary conditions of the Fubini instanton are $\frac{d\Phi}{d\eta}|_{\eta=0}=0$ and $\Phi|_{\eta=\infty}=0$ as in Ref.\ \cite{fubi01}. The first condition is for the solution being regular at the origin. The second condition is for the requirement to describe the outside state of the solution, i.e.\ the initial background configuration. This makes the decay probability finite. The solutions exist for an arbitrary $\Phi_o$ due to the scale invariance. We can interpret the equation of motion as follows: The particle starts with the zero velocity at $\Phi=\Phi_o$ and rolls down to the bottom in the inverted potential. Finally, the particle stops at $\Phi=0$ at $\eta=\infty$ without any oscillation. In the presence of gravity, we should impose two additional boundary conditions for $\rho(\eta)$. The geometry is noncompact for flat or AdS space and compact for dS space which reads us for the convenience to choose the different type of boundary conditions at $\eta=0$ and $\eta=\eta_{max}$. The $\eta_{max}$ is infinite for flat and AdS spaces while is is finite for dS space.

For flat or AdS space, we can impose boundary conditions as follows \cite{lllo}:
\begin{equation}
\rho|_{\eta=0}=0, \,\,\,\, \frac{d\rho}{d\eta}\Big|_{\eta=0}=1, \,\,\,\,
\frac{d\Phi}{d\eta}\Big|_{\eta=0}=0, \,\,\,\, {\rm and}\,\,\,\,
\Phi|_{\eta=\eta_{max}}=0 \,.\label{Eq.ourbc-2}
\end{equation}
The first condition is for a geodesically complete space. The second condition stems from Eq.\ (\ref{eom3}).
For dS space, we can impose boundary conditions as follows:
\begin{equation}
\rho|_{\eta=0}=0 ,
\,\,\,\, \rho|_{\eta =\eta_{max}} =0, \,\,\,\,
\frac{d\Phi}{d\eta}\Big|_{\eta=0}=0, \,\,\,\, {\rm and}\,\,\,\,
\frac{d\Phi}{d\eta}\Big|_{\eta =\eta_{max}} = 0. \label{Eq.ourbc-3}
\end{equation}
Analytic solutions are not known in the presence of gravity, hence we employ the numerical computation. To solve the Euclidean field equations (\ref{eom2}) and (\ref{eom3}) numerically, we make dimensionless variables as in Ref.\ \cite{bwdd}. In this procedure, the parameter $\kappa$ corresponds to the ratio between Planck mass and the mass scale in the theory. In what follows, we employ dimensionless variables without a tilde. We choose the initial values of
$\Phi(\eta_{\mathrm{initial}})$, $\Phi'(\eta_{\mathrm{initial}})$, $\rho(\eta_{\mathrm{initial}})$, and
$\rho'(\eta_{\mathrm{initial}})$ at $\eta_{\mathrm{initial}}= 0+\epsilon$ for $\epsilon \ll 1$ as follows:
\begin{eqnarray}
\Phi(\epsilon) &\simeq& \Phi_{o} - \frac{\epsilon^2}{8}\Phi^3_o +  \cdot\cdot\cdot \,,  \nonumber \\
\Phi'(\epsilon) &\simeq& - \frac{\epsilon}{4}\Phi^3_o + \cdot\cdot\cdot \,,  \label{nbcon2} \\
\rho(\epsilon) &\simeq& \epsilon + \cdot\cdot\cdot \,,  \nonumber \\
\rho'(\epsilon) &\simeq& 1 + \cdot\cdot\cdot \,. \nonumber
\end{eqnarray}
The initial value of $\Phi'$ is taken to be positive in the present paper. Once we specify the initial position $\Phi_o$, then all other quantities can be exactly determined from Eq.\ (\ref{nbcon2}). To diminish numerical errors Taylor expansion with higher precisions was considered in Ref.\ \cite{hprecisions}.

To get the tunneling probability, we only need to consider the bulk part of the Euclidean action, since the contribution from the YGH boundary term between the bounce solution and the background cancels out each other. The bulk action is evaluated as follows:
\begin{equation}
S_E= \int_{\mathcal M} \sqrt{g_{\mathrm{E}}} d^4 x_{\mathrm{E}}
\left[ -\frac{R_\mathrm{E}}{2\kappa} +\frac{1}{2}\Phi'^2 +U \right]
=  2\pi^2 \int \rho^3 d\eta [-U]\,, \label{euclac}
\end{equation}
where $R_{\mathrm{E}} =6[1/\rho^2 - \rho'^2/\rho^2 - \rho''/\rho]$. We used Eqs.\ (\ref{eom2}) and (\ref{eom3}) to get the last expression. We define the ``Euclidean action density" as $E_{\xi}(\eta)=2\pi^2\rho^3\xi$. Then Eq.\ (\ref{euclac}) implies the density $\xi=-U$. Now we evaluate the action difference. If one use the variable $d\eta$ the upper bounds of $\eta$ for both actions have the different value. In other words, $\eta_{bs}(\bar\rho)$ is different from $\eta_{bg}(\bar\rho)$, in which $\bar\rho$ is the radius of a bounce solution. Hence, we change the variable $d\eta$ into $d\rho$. As a result, the action difference $B$ is written as follows:
\begin{equation}
B= 2\pi^2 \int^{\rho_{max}}_{0} \left[\frac{\rho^3 d\rho [-U(bs)]}{\sqrt{1+\frac{\kappa\rho^2}{3}[\frac{1}{2}\Phi'^2-U(bs)] }} -  \frac{\rho^3 d\rho [-U(bg)]}{\sqrt{1+\frac{\kappa\rho^2}{3}[-U(bg)] }} \right]\,, \label{semiexpob}
\end{equation}
where $\rho_{max}$ is the maximum value of the radius $\rho$. For the case of dS space, $\rho_{max}$ is the radius of $4$-sphere, which is finite. Thus, the probability is guaranteed to be finite. For the cases of AdS space and flat space, the $\rho_{max}$ is infinite. Therefore we should check the probability more carefully. In the next section and Appendix we will straightforwardly compute the action difference according to Eq.\ (\ref{semiexpob}).

\section{Various types of numerical solutions \label{Sec.3}}

In this section, we numerically solve the coupled equations of the scalar field and gravity. We concentrate on various types of numerical solutions both in AdS and dS spaces. We will show that the solutions can be classified by the number of oscillations.

\subsection{Computational methods \label{Sec.3.1}}

In order to solve the coupled equations of motion numerically, we employ the fourth-order Runge-Kutta method with the Euclidean evolution parameter step size of $10^{-7}$. Three quantities, $\kappa$, $U_o$, and $\Phi_o$, are chosen as numerical parameters. They correspond to the reduced Newtonian gravitational constant, the maximum value of the potential when the initial background state is located at $\Phi=0$, and the initial value of a scalar field in numerical computation, respectively. The parameter $U_o$ is related to the cosmological constant as $\Lambda = \kappa U_o$. The background space is dS, flat and AdS depending on the values of $U_o$. The maximum value of the evolution parameter $\eta$, the Euclidean time, is also determined by the values of $U_o$. It is finite for dS space, while it is infinite for AdS and flat space. Consequentially, the types of solutions depend on the values of $U_o$. In the absence of gravity with $\kappa=0$, there is only one parameter $\Phi_o$, regardless of $U_o$. For an arbitrary value of $\Phi_o$, there are bounce solutions without oscillation. The initial value $\Phi_o$ is related to the size of the instanton.

\subsection{Numerical results} \label{Sec.3.2}

\begin{figure}[t]
  \centering
  \includegraphics[width=0.65\textwidth]{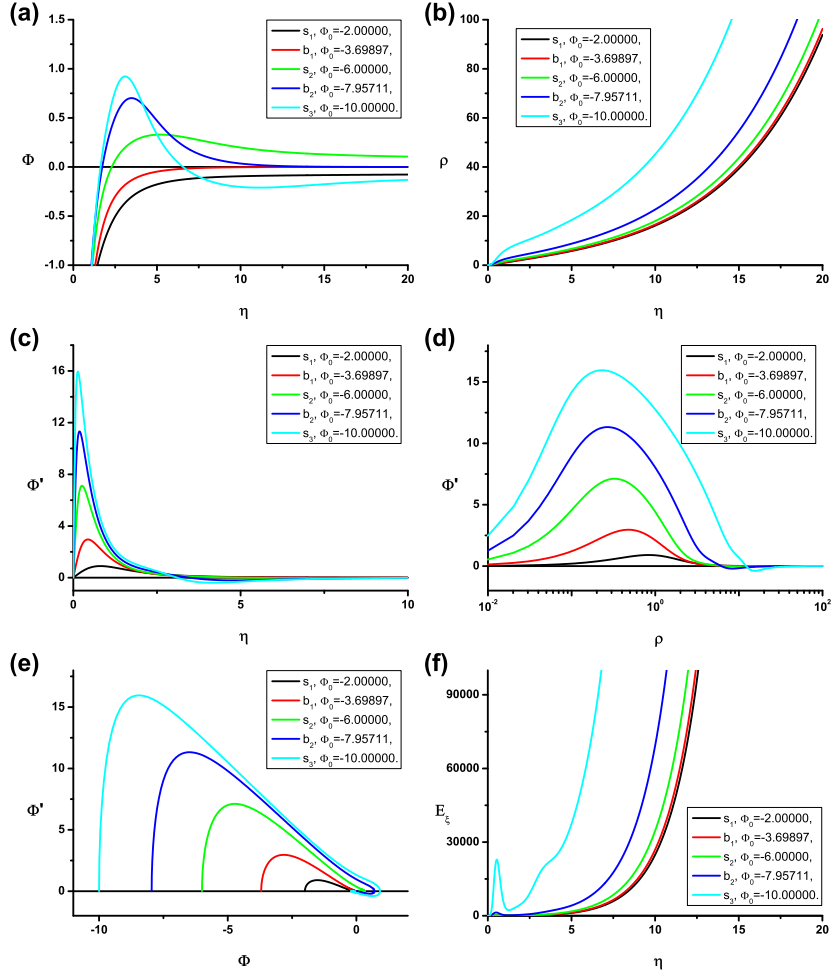}
  \caption{\footnotesize{(color online). (a) Numerical solutions for $\Phi$, (b) solutions for $\rho$, (c) the variation of $\Phi'$ with respect to $\eta$, (d) the variation of $\Phi'$ with respect to $\rho$, (e) phase diagram of $\Phi'$ versus $\Phi$, and (f) Euclidean action density $E_{\xi}$ evaluated at constant $\eta$ in AdS space. We take $\kappa=0.30$ and $U_o = -0.30$.}}
  \label{Fig.AdS_solutions}
\end{figure}

We first solve Eqs.\ (\ref{eom2}) and (\ref{eom3}) with the boundary conditions (\ref{Eq.ourbc-2}) in AdS space. We take $\kappa=0.30$ and $U_o = -0.30$. The solution exists for any initial value $\Phi_o$. The numerical solutions for $\Phi$ and $\rho$ in AdS space are shown in Fig.\ \ref{Fig.AdS_solutions} for five different initial values $\Phi_o$. That is, $-2.00000$ presented with the black line, $-3.69897$ with the red line, $-6.00000$ with the green line, $-7.95711$ with the blue line, and $-10.00000$ with the sky blue line, respectively. Figure \ref{Fig.AdS_solutions}(a) presents the solutions for $\Phi$ with respect to $\eta$. The profile $\Phi(\eta)$ drops from $\Phi_o$ at $\eta=0$ and asymptotically approaches to $0$ at $\eta\rightarrow \infty$. In each figure, we denote the oscillating solutions as $s_i$, where the index $i$ represents the number of oscillations. An ordinary bounce solution ``oscillates'' one, i.e., $i=1$. The black, green and sky blue lines correspond to the solutions with the oscillation $1$, $2$, and $3$ times, respectively. We denote the oscillating solution at the boundary between $s_j$ and $s_{j+1}$ solutions as $b_j$. In other words, $b_j$ is the marginal solution of $s_j$ solutions. The red and blue lines correspond to the marginal solutions with the oscillation $1$ and $2$ times, respectively. There is the tendency that the number of oscillations is decreased as the value of $\Phi_o$ is decreased. This solution has the thick wall separating the true vacuum consisted of only one point at the center of the solution from the outside false vacuum state. Figure \ref{Fig.AdS_solutions}(b) presents the solutions for $\rho$. The curves move upwards with increasing value of $|\Phi_o|$. The general behavior of the numerical solution can be easily understood if one thinks of the shape of the solution in a fixed AdS space as $\rho = \sqrt{\frac{3}{\Lambda}}\sinh \sqrt{\frac{\Lambda}{3}}\eta$. Figure \ref{Fig.AdS_solutions}(c) presents $\Phi'$ with respect to $\eta$. The value of $\Phi'$ has the maximum value around $\eta\leq1$. This means that the field rolls down the inverted potential, its velocity has the maximum value at the very early state around $\eta\leq1$, and the velocity is decreased as $\eta$ goes to infinity. In the usual bubble solutions, the maximum velocity occurs in the middle of the bubble wall. Therefore, one could consider the location as the size of the instanton. For the Fubini instanton whose shape is highly asymmetric, it is not clear how to determine the size of the instanton. Figure \ref{Fig.AdS_solutions}(d) presents the solutions for $\Phi'$ with respect to $\rho$. Figure \ref{Fig.AdS_solutions}(e) presents $\Phi'$ versus $\Phi$. Each trajectory begins with zero velocity $\Phi'=0$. The velocity rapidly increases to the maximum and then decreases linearly up to the turning point as shown in Ref.\ \cite{bwdd}. Figure \ref{Fig.AdS_solutions}(f) presents the Euclidean action density $E_{\xi}(\eta)$ whose integration over $\eta$ gives the Euclidean action $S_E$. The first peak is due to nontrivial contributions from the potential and kinetic energy. The increase of the Euclidean action density with respect to $\eta$ is due to the increase of $\rho$ in AdS space.

We now compute the action difference for the solutions in Fig.\ \ref{Fig.AdS_solutions}. We examine whether or not the scalar field has the exponentially decaying property in the asymptotic region to give the action difference finite.

First, we carry out the action integral in the range of $0\leqq \rho \leqq 10^{11}$ numerically. We straightforwardly compute the action difference $B$ using Eq.\ (\ref{semiexpob}). As shown in Appendix, the exponent $B$ diverges for the solutions $s_j$, while finite for the marginal solutions $b_j$. We summarize the exponent $B$ for two marginal solutions in Table \ref{Tab.AdS_solution}.
\begin{table}[t]
	\centering	
    \newcolumntype{A}{>{\centering\arraybackslash} m{2cm} }	
    \newcolumntype{B}{>{\centering\arraybackslash} m{3cm} }		
    \begin{tabular}{A A A B B m{0cm}}		
       \hline		
       Type & $\kappa$ & $U_o$ & $\Phi_o$ & $B$ & \\ [1ex]		
       \hline		
       $b_1$ & \multirow{2}{*}[-0.5ex]{$0.30$} & \multirow{2}{*}[-0.5ex]{$-0.30$} & $-3.69897$ & $37.6506$ & \\ [1ex]	
       $b_2$ & & & $-7.95711$ & $579.9922$ & \\ [1ex]
       \hline	
    \end{tabular}	
    \caption{\footnotesize{Parameter choices and exponent $B$s for the marginal solutions $b_j$ plotted in Fig.\ (\ref{Fig.AdS_solutions}).}}
    \label{Tab.AdS_solution}
\end{table}
We comment that there are hopes to regularize Euclidean actions for $s_j$ using the no-boundary regulator \cite{thja}. This method is also worthwhile to investigate further for quartic potentials, while we only have restricted for the quadratic potentials in \cite{thja}. As a different check, we postpone this issue for future papers.

\begin{figure}
  \centering
  \includegraphics[width=0.8\textwidth]{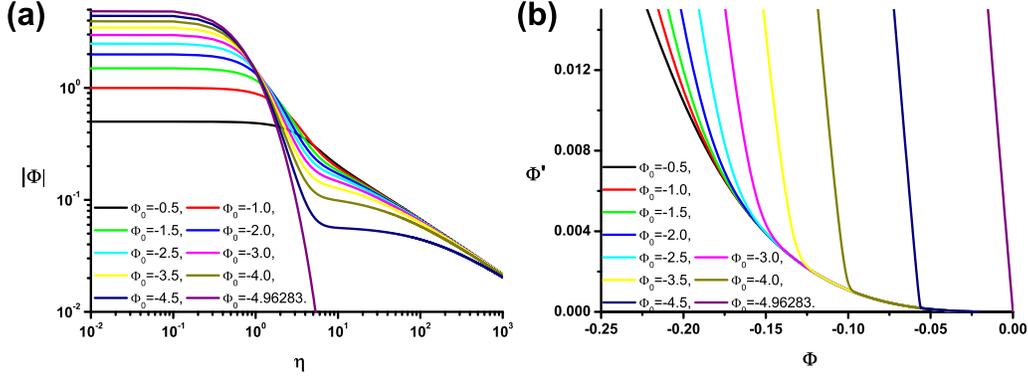}
  \caption{(color online). (a) Log-log scale plot of $\Phi$ versus $\eta$ and (b) the phase diagram of $\Phi'$ versus $\Phi$ for the nine $s_1$ solutions with the initial position starting from $\Phi_o=-0.5$ to $\Phi_o=-4.0$ and one $b_1$ solution (purple line) with $\Phi_o=-4.96283$. We choose $\kappa=0.30$ and $U_o=-1.00$.}
  \label{Fig.AdS_converge_test}
\end{figure}

Second, we present the qualitative argument for evaluating the exponent. The potential near the tachyonic top can be expanded as follows:
\begin{equation}
U=U_o - \frac{1}{2}m^2\Phi^2 - \frac{\lambda}{4}\Phi^4 \cdot\cdot\cdot  .
\end{equation}
Around the asymptotic values of $\eta$, the approximate behavior of the scalar field becomes
\begin{equation}
\Phi \simeq A_1 \exp \left[ \left(-\frac{3}{2} + \sqrt{\frac{9}{4} - \frac{m^2}{H^2}} \right)H\eta \right] +A_2 \exp \left[ \left(-\frac{3}{2} - \sqrt{\frac{9}{4} - \frac{m^2}{H^2}} \right) H\eta \right],
\end{equation}
where $H^2\simeq -\kappa U_o/3$ and $A_{1, 2}$ are constants \cite{llry5}. According to Ref.\ \cite{kamso}, the condition of the converging action is $A_1=0$. If the tachyonic top has only the quartic term, then approximately
\begin{equation}
\Phi \simeq A_1 + A_2 \exp[-3H\eta].
\end{equation}
For asymptotic AdS space, $A_1$ should be controlled to be zero. This suggests that the tunneling probability may be finite. However we need to numerically confirm the behavior of the scalar field in the asymptotic region.

Last, therefore, we examine the scalar field in more detail how fast it approaches to zero in the asymptotic region. Figure \ref{Fig.AdS_converge_test} presents the log-log scale plot of $\Phi$ versus $\eta$, and the phase diagram of $\Phi'$ versus $\Phi$ in the asymptotic region. We take $\kappa=0.30$ and $U_o=-1.00$. In this figure the nine $s_1$ solutions and one $b_1$ solution (purple line) are plotted. In Fig.\ \ref{Fig.AdS_converge_test}(a), we can see that all $s_1$ solutions whose initial positions are starting from $\Phi_o=-0.5$ to $\Phi_o=-4.5$ converge to a certain linear line. Using this property, we approximate a scalar field for the late-time behavior as
\begin{equation}
\log|\Phi| \ =\ -E\log \eta + F,
\label{Eq.AdS_approx1}
\end{equation}
where the $E$ and $F$ are positive constants. The above equation is reduced to
\begin{equation}
|\Phi| \ =\ F' \eta^{-A},
\label{Eq.AdS_approx2}
\end{equation}
where $F'=e^F$ which is a constant. Therefore, we can see that the $s_1$ solutions with above late-time behaviors obey Eq.\ (\ref{Eq.AdS_approx2}) and the scalar field goes to zero when $\eta$ goes to infinity. This asymptotic power behavior makes the action infinity. However, the late-time behavior of the $b_1$ solution (purple line, $\Phi_o=-4.96283$) is totally different from those of other solutions. The slope of the line tends to increase to approximately $\log\eta = {\rm constant}$ as $\eta$ increases. This suggests that the $b_1$ solution may reach $\Phi=0$ and $\Phi'=0$ within a finite time. In Fig.\ \ref{Fig.AdS_converge_test}(b), $s_1$ solutions are shown to approach to the asymptotic curve. In other words, all $s_1$ solutions converge to the curve exponentially. Thus, $s_1$ solutions converge to the origin point when $\eta$ goes to infinity. However, the $b_1$ solution falls down directly at $\Phi=0$ and $\Phi'=0$. This analysis shows that the oscillations are finished before the scalar field reaches zero. Thus, if the scalar field shows the late-time asymptotic behavior, we can stop at that finite $\eta$ to count the number of oscillations. The similar analysis can be applied for $i> 1$. This is the reason why we are safe to cut the Euclidean time when we count the number of oscillations.

In summary of this part, the marginal solutions $b_j$ are shown to have a finite probability, while $s_j$ solutions have a vanishing probability. In the Appendix, the behavior of $\Phi$ and the exponent $B$ are numerically examined in more detail.

\begin{figure}[t]
  \centering
  \includegraphics[width=0.9\textwidth]{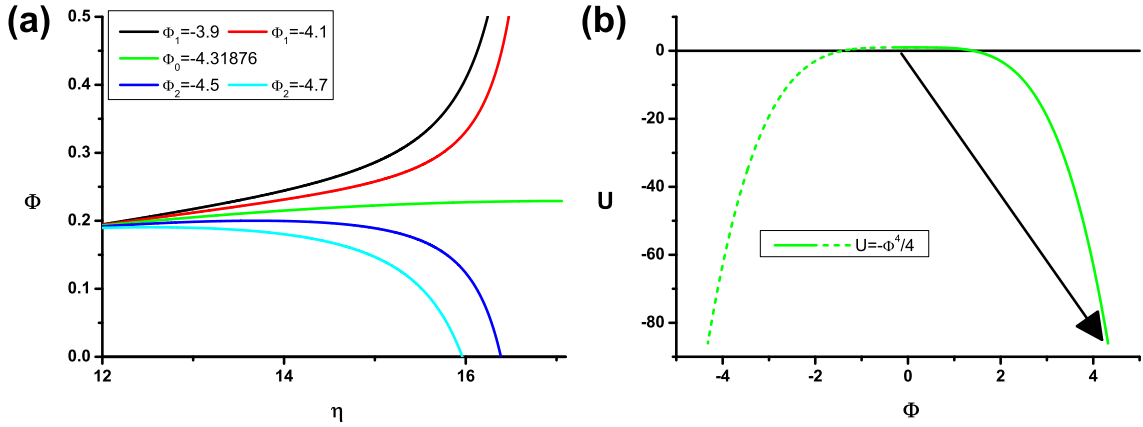}
  \caption{\footnotesize{(color online). The figure represents the behaviors of scalar fields with different initial values which is to describe the undershoot-overshoot procedure for the case in dS space. We take $\kappa=0.10$ and $U_o=1.00$.}}
  \label{Fig.dS_divergence}
\end{figure}

Now we are going to turn to the solutions in dS space. Unlike the solutions in AdS space, the solution exists only for some values $\Phi_o$ in dS space. Hence, we adopt the so-called undershoot-overshoot procedure \cite{col002} to get the solutions in dS space. Suppose we have two initial values $\Phi_1$ and $\Phi_2$ which are diverging to positive and negative infinity, respectively. Then, we expect that there is the initial value $\Phi_o$ between $\Phi_1$ and $\Phi_2$ which does not give the divergence to infinity. We choose $\kappa=0.10$ and $U_o=1.00$. Figure \ref{Fig.dS_divergence} presents the behaviors of scalar fields with different initial values. Since the system is invariant under $\Phi \rightarrow -\Phi$, the only one signature of $\Phi_o$ can be chosen without loss of generality. In the numerical computations, only specific parameter value $\Phi_o= -4.31876$ can satisfy given boundary conditions Eq.\ (\ref{Eq.ourbc-3}) as shown in Fig.\ \ref{Fig.dS_divergence}. Other values are diverging to the positive or negative infinity at $\eta_{max}$. Note that the instanton solution stops near the point $\Phi=0$ rather than at $\Phi=0$ [see the green line in Fig.\ \ref{Fig.dS_divergence}(a)]. This means that the tunneling occurs from near the top to a certain state in dS space. The corresponding tunneling process is drawn in Fig.\ \ref{Fig.dS_divergence}(b).

We note that the Euclidean compact dS space is invariant under the $Z_2$ transformation $\eta \rightarrow \eta_{\text{max}}-\eta$. For this reason, there exist solutions with $Z_2$ symmetry of even or odd parity for some specific values $\Phi_o$ as shown in Ref.\ \cite{bwdd}.

\begin{figure}[t]
  \centering
  \includegraphics[width=0.7\textwidth]{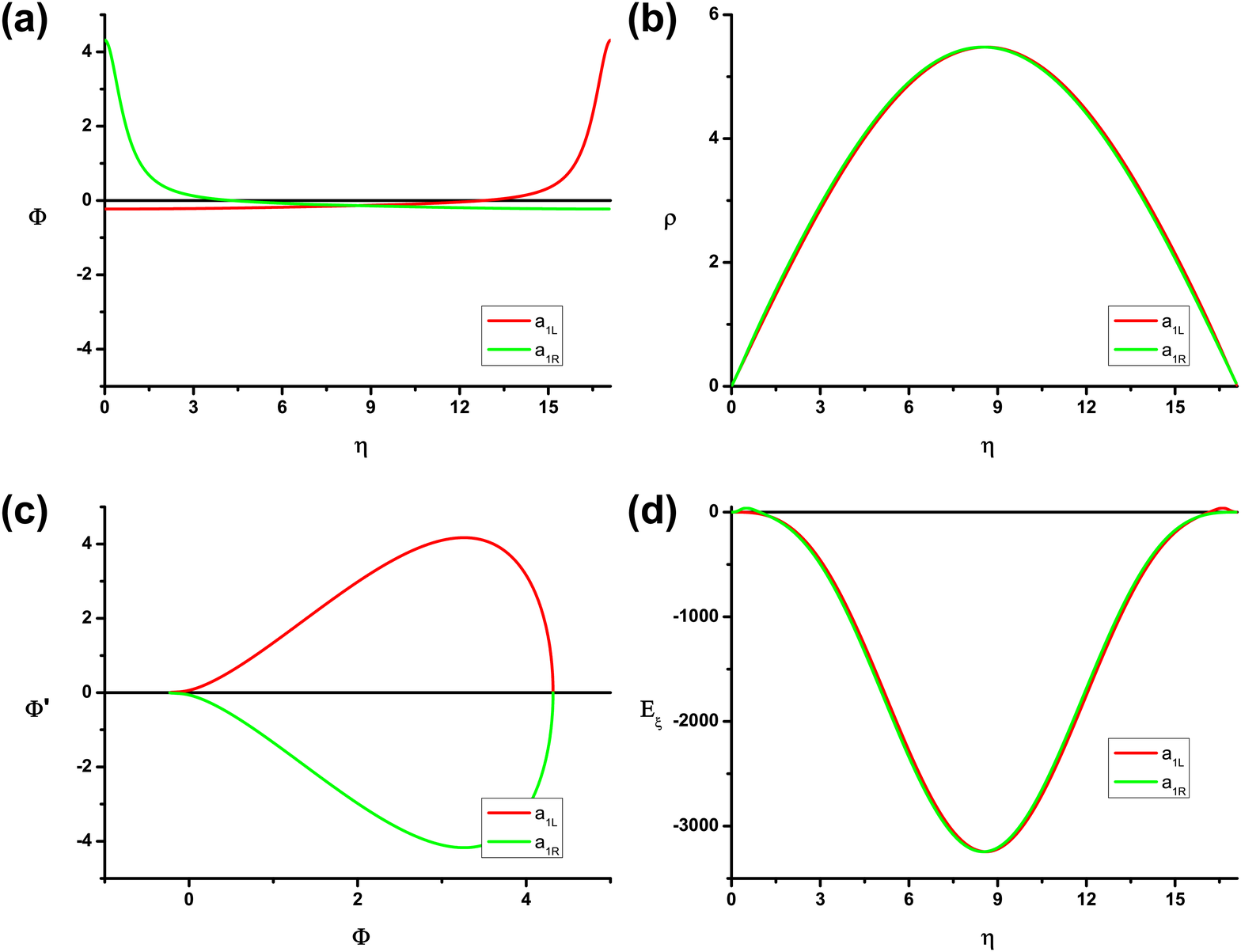}\\
  \caption{\footnotesize{(color online). (a) Numerical solutions for $\Phi$, (b) solutions for $\rho$, (c) phase diagram of $\Phi'$ versus $\Phi$, and (d) Euclidean action density $E_{\xi}$ evaluated at constant $\eta$ of $Z_2$ asymmetric cases in dS space. We choose $\kappa=0.10$ and $U_o=1.00$.}}
  \label{Fig.dS_asymmetric1}
\end{figure}

\begin{figure}[t]
  \centering
  \includegraphics[width=0.7\textwidth]{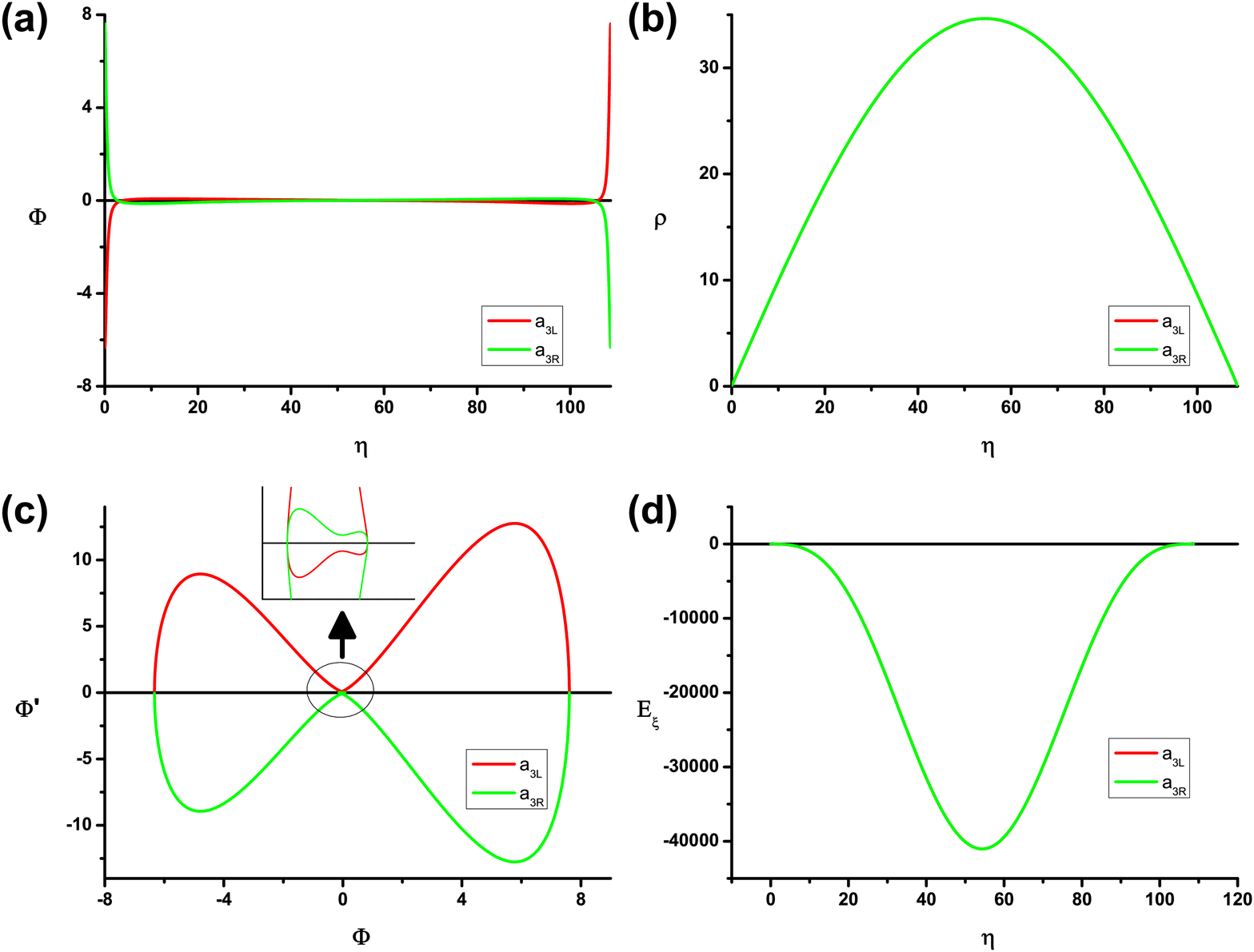}\\
  \caption{\footnotesize{(color online). (a) Numerical solutions for $\Phi$, (b) solutions for $\rho$, (c) phase diagram of $\Phi'$ versus $\Phi$, and (d) Euclidean action density $E_{\xi}$ evaluated at constant $\eta$ of $Z_2$ asymmetric cases in dS space. We choose $\kappa=0.05$ and $U_o=0.05$.}}
  \label{Fig.dS_asymmetric2}
\end{figure}

In addition, there can be $Z_2$-asymmetric solutions for other values $\Phi_o$ as shown in Figs.\ \ref{Fig.dS_asymmetric1} and \ref{Fig.dS_asymmetric2}. These solutions are denoted as $a_{iL}$ (green line) and $a_{iR}$ (red line), which are connected by $Z_2$ transformation. The subscript index $i$ is the number of oscillation. $L$ and $R$ mean $\Phi_o < 0$ and $\Phi_o > 0$, respectively.

Figure \ref{Fig.dS_asymmetric1}(a) presents the numerical solutions for $\Phi$ with the number of oscillations $i$ equal to $1$. The $a_{1L}$ (green line) starts at $4.31876$ with zero velocity, is decreasing, monotonically passing the point $\Phi=0$, and arrives at $-0.22912$. The $a_{1R}$ (red line) starts at $-0.22912$ with zero velocity, is monotonically passing the point $\Phi=0$, increasing due to the antidamping term, and arrives at $4.31876$. The Fubini instanton solution stops near the point $\Phi=0$ in dS space as mentioned before. To see that this is consistent with equations of motion, we can estimate numerically the magnitudes in the asymptotic region as $\frac{dU}{d\Phi} \sim - \Phi^3_{fs}$ and $\frac{3\rho'}{\rho}\Phi' \sim - \frac{3}{4}\Phi^3_{fs}$. The choice of $\Phi'' \sim - \frac{1}{4}\Phi^3_{fs}$ satisfies the equations of motion.
For the cases of flat and AdS spaces, the equations of motion are trivially satisfied since $\Phi_{fs}$ goes to zero.
Figure \ref{Fig.dS_asymmetric1}(b) shows the numerical solutions for $\rho$. Figure \ref{Fig.dS_asymmetric1}(c) presents the phase diagram of $\Phi'$ versus $\Phi$. The first rising stage of the red curve satisfies $d\Phi'/d\Phi=c$, i.e.\ a positive constant, the second stage $d\Phi'/d\Phi=0$, and the third stage $\Phi'\simeq - \sqrt{\frac{\lambda}{2}(\Phi_o^4 - \Phi^4)}$ \cite{bwdd}. Figure \ref{Fig.dS_asymmetric1}(d) presents the Euclidean action density $E_{\xi}(\eta)$.

Figure \ref{Fig.dS_asymmetric2}(a) presents $\Phi$ with $i$ equal to $3$. The $a_{3L}$ (green line) starting at $7.60997$ with zero velocity is decreasing, then oscillates passing the point zero ($\Phi=0$) three times, decreasing, and arrives at $-6.33115$. The $a_{3R}$ (red line) starting at $-6.33115$ with zero velocity is increasing, oscillates passing the point zero ($\Phi=0$) three times, increasing, and arrives at $7.60997$. Figure \ref{Fig.dS_asymmetric2}(b) presents the numerical solutions for $\rho$. Figure \ref{Fig.dS_asymmetric2}(c) presents the phase diagram of $\Phi'$ versus $\Phi$. The first rising stage of the red curve starting at the most left satisfies $\Phi'\simeq  \sqrt{\frac{\lambda}{2}(\Phi_o^4 - \Phi^4)}$, the second stage $d\Phi'/d\Phi=0$, the third stage $d\Phi'/d\Phi=-c$, i.e. a negative constant, the fourth stage the behavior near the origin point, the fifth stage $d\Phi'/d\Phi=c$, the sixth stage $d\Phi'/d\Phi=0$, and the seventh stage $\Phi'\simeq - \sqrt{\frac{\lambda}{2}(\Phi_o^4 - \Phi^4)}$. The upper box in the same figure shows the magnification of the small region representing behavior of the curves, which shows each curve is passing the point zero ($\Phi=0$) three times. Figure \ref{Fig.dS_asymmetric2}(d) presents the Euclidean action density $E_{\xi}(\eta)$. The green and red lines are overlapping in Figs.\ \ref{Fig.dS_asymmetric2}(b) and \ref{Fig.dS_asymmetric2}(d). We summarize the parameter choices and probabilities of several $Z_2$-asymmetric solutions in Table \ref{Tab.dS_asymmetric}.

\begin{table}[H]
	\centering
	\newcolumntype{A}{>{\centering\arraybackslash} m{1.5cm} }
	\newcolumntype{B}{>{\centering\arraybackslash} m{2cm} }
	\begin{tabular}{A A A B B B A m{0cm}}
		\hline
		Type & $\kappa$ & $U_o$ & $\Phi_o$ & $S^{cs}$ & $S^{bg}$ & $B$ & \\ [1ex]
		\hline
		$a_{1L}$ & \multirow{2}{*}[-0.5ex]{$0.10$} & \multirow{2}{*}[-0.5ex]{$1.00$} & $-0.22912$ & $-23654$ & \multirow{2}{*}{$-23687$} & $33$ & \\ [1ex]
		$a_{1R}$ & & & $4.31876$ & $-23654$ &  & $33$ & \\ [1ex]
		\hline
		$a_{3L}$ & \multirow{2}{*}[-0.5ex]{$0.05$} & \multirow{2}{*}[-0.5ex]{$0.05$} & $-6.33115$ & $-189486$ & \multirow{2}{*}{$-189496$} & $10$ & \\ [1ex]
		$a_{3R}$ & & & $7.60997$ & $-189486$ &  & $10$ & \\ [1ex]
		\hline
	\end{tabular}
	\caption{\footnotesize{Parameter choices and probabilities of $Z_2$-asymmetric solutions}}
	\label{Tab.dS_asymmetric}
\end{table}

\section{Parametric phase diagram \label{Sec.4}}

In this section, we construct the parametric phase diagram of various solutions with respect to the parameters $\kappa$, $U_o$ and $\Phi_o$. There are several types of solutions including oscillating solutions as we discussed in Sec.\ \ref{Sec.3}. We investigate the properties of those solutions.

\subsection{Computational methods \label{Sec.4.1}}

\begin{figure}
  \centering
  \includegraphics[width=0.8\textwidth]{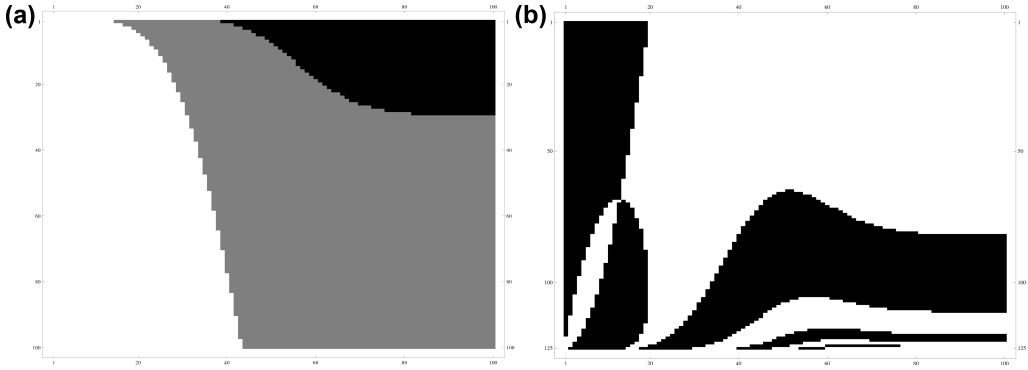}\\
  \caption{\footnotesize{Matrix plots in (a) AdS space with white, gray and black colors which correspond to the different number of oscillations such as 1, 2, and 3, respectively, and in (b) dS space with white and black colors which correspond to the direction of divergence such as negative and positive, respectively. We choose $\kappa=0.30$.}}
  \label{Fig.matrixplot}
\end{figure}

We employ the matrix plot that gives a visual representation of the values of elements in a matrix. Figure \ref{Fig.matrixplot} presents the matrix plot for visual representation of the oscillating solutions in AdS and dS spaces. We choose $\kappa=0.30$.
We divide into two parametric phase diagrams depending on the values of $U_o$. The $X$-axis in each plot corresponds to $\Phi_o$, which is from $0$ to $-14.00$, and the $Y$-axis corresponds to $U_o$. Because this process can show only approximated points of solutions, we should gather more dense data in all regions.

Figure \ref{Fig.matrixplot}(a) presents the parametric phase diagram in AdS space. The $Y$-axis is from $0$ to $-2.00$ where the direction is downward. We first divide the $X$-axis and the $Y$-axis as 100 pieces. In this case, $\eta$ grows up to infinity however we are safe to count the number of oscillations during the finite Euclidean time as we discussed at Sec.\ \ref{Sec.3.1}. We count the number of oscillations for every point in the phase diagram. In general, this process can only recognize the $s_i$ solutions. Thus, we need to take a more elaborated approach to find the $b_j$ solutions and get the data more dense rather than those in Fig.\ \ref{Fig.matrixplot}(a). For AdS space, there exist an infinite number of solutions occupying the area composed of two parameters $U_o$ and $\Phi_o$ with given value of $\kappa$ as shown in Fig.\ \ref{Fig.matrixplot}(a).

Similarly with the previous figure, Fig.\ \ref{Fig.matrixplot}(b) presents the parametric phase diagram in dS space. The $Y$-axis is from $0$ to $2.50$ where the direction is upward. We divide the $X$-axis as $100$ pieces and the $Y$-axis as $125$ pieces. $\eta$ is finite as $\eta_{max}$. We check the divergence at all points using the shooting method, i.e.\ undershoot-overshoot procedure. The result is illustrated in Fig.\ \ref{Fig.matrixplot}(b). The oscillating solutions with both $Z_2$-symmetric and asymmetric solutions are included. There exist the solutions only on the curve.

\subsection{Numerical results \label{Sec.4.2}}

\begin{figure}[t]
  \centering
  \includegraphics[width=0.8\textwidth]{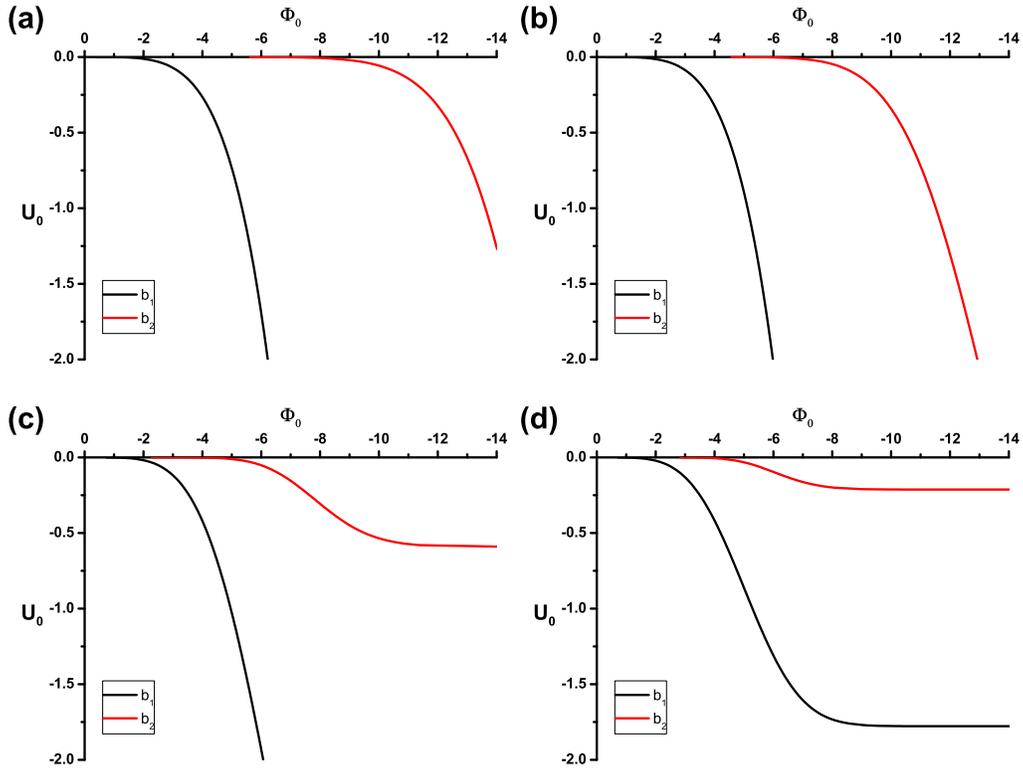}\\
  \caption{\footnotesize{(color online). Parametric phase diagram in AdS space with (a) $\kappa=0.05$, (b) $\kappa=0.10$, (c) $\kappa=0.30$, and (d) $\kappa=0.50$, respectively.}}
  \label{Fig.AdS_phase}
\end{figure}

We consider four cases for different values of $\kappa$. We plot the parametric phase diagrams of $U_o$ and $\Phi_o$ in AdS and dS spaces.

We choose $\kappa=0.05$, $0.10$, $0.30$, and $0.50$. Figure \ref{Fig.AdS_phase} presents the phase diagrams in AdS space.
In this figure, we consider only up to $i$ equal to $3$ oscillating solutions. There are solutions with oscillation more than $3$ near the $U_o$ equal to $0$. However, we do not display these solutions because of the numerical difficulty. We take $\kappa=0.05$, $0.10$, $0.30$, and $0.50$ for (a), (b), (c), and (d), respectively. $U_o$ is from $-2.0\times10^{-4}$ to $-2.00$ and $\Phi_o$ is from $0$ to $-14.00$. The black lines correspond to the $b_1$ solutions and the red lines the $b_2$ solutions for the given range of $U_o$'s. The points out of lines are $s_i$ solutions with the oscillation $i$ times. Consequentially, the solutions fill up the whole area composed of two parameters, $U_o$ and $\Phi_o$.

We can observe the behavior of the solutions with respect to the parameter values in Fig.\ \ref{Fig.AdS_phase}. In general the number of oscillations is decreased as the value of $\Phi_o$ is decreased. Let us fix the value $\Phi_o$ in each figure. The higher number of oscillating solutions could exist as the value of $|U_o|$ is decreased. As can be seen from four figures, $b_j$ solutions move to the left and upward as $\kappa$ is increased. From this behavior of $b_j$ solution, we can see that there is the minimum value of $U_o$ which does not allow the higher number of oscillations. For example, if we fix $U_o=-1.00$ then the solution can oscillate up to $i$ equal to $3$. All $b_j$ solutions have similar property. Moreover, for any number of oscillations the initial values $\Phi_o$s which give the $b_j$ solutions are approaching to zero as $U_o$ goes to zero. To see this behavior more clearly, we examine the behavior using the log-log scale plot of the absolute values of $U_o$ and $\Phi_o$.

\begin{figure}[t]
  \centering
  \includegraphics[width=0.7\textwidth]{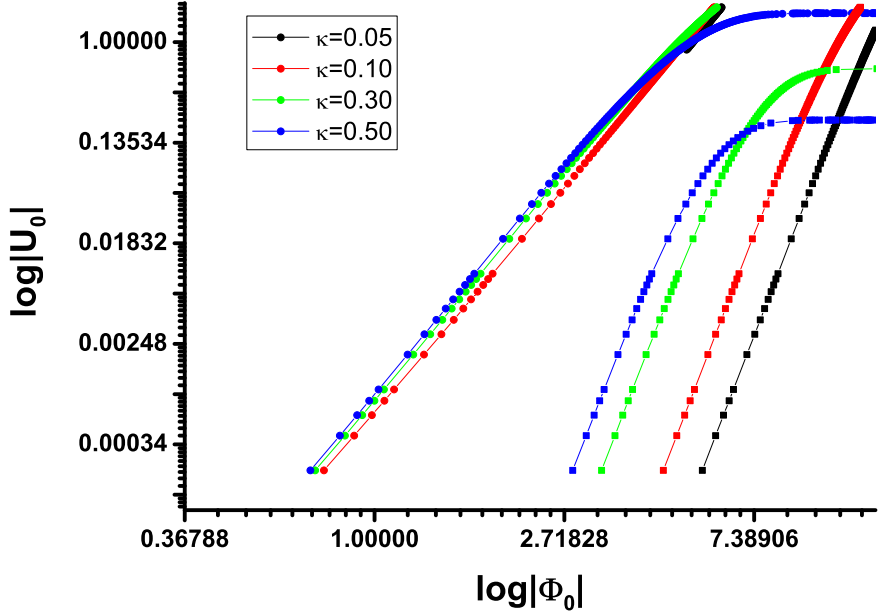}\\
  \caption{\footnotesize{(color online). Parametric phase diagram of the log-log scale of $|U_o|$ and $\Phi_o$ with four different values of $\kappa$ in AdS space.}}
  \label{Fig.AdS_phase_boundary}
\end{figure}

Figure \ref{Fig.AdS_phase_boundary} shows the log-log scale plot with same marginal solutions used in Fig.\ \ref{Fig.AdS_phase}. The left four curves correspond to $b_1$, while the right four curves correspond to $b_2$. Because $b_1$ and $b_2$ solutions with respect to $U_o$ and $\Phi_o$ are linear in the vicinity of $U_o=0$ and $\Phi_o=0$, each of the solutions has a relation between these two parameters as follows:
\begin{equation}
\label{Eq.dS_approx}
\log|U_o|\ =\ E \log |\Phi_o| + F \quad \Rightarrow \quad U_o=\pm e^{F}|\Phi_o|^{E},
\end{equation}
where $E$ and $F$ are positive constant. We take the $-$ sign for our case and the other sign for $\Phi_o>0$. We can easily see that the $b_1$ solution and the $b_2$ solution approach to the point $U_o=0$ and $\Phi_o=0$ from Eq.\ (\ref{Eq.dS_approx}). From this numerical result, we can expect that all $b_j$ solutions behave like the $b_1$ solution.

\begin{figure}[t]
  \centering
  \includegraphics[width=0.8\textwidth]{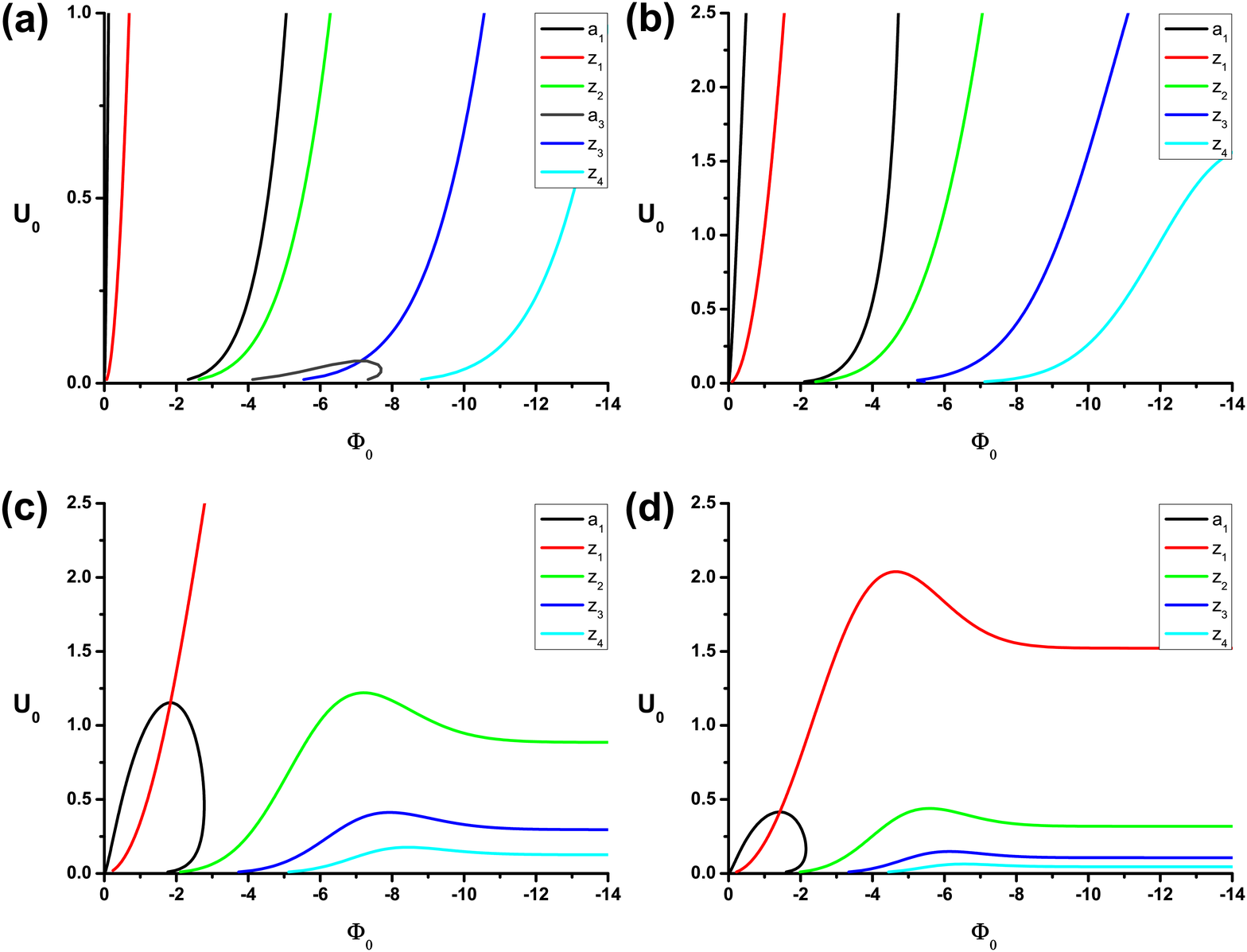}\\
  \caption{\footnotesize{(color online). Parametric phase diagrams in dS space with (a) $\kappa=0.05$, (b) $\kappa=0.10$, (c) $\kappa=0.30$, and (d) $\kappa=0.50$, respectively.}}
  \label{Fig.dS_phase}
\end{figure}

Figure \ref{Fig.dS_phase} presents parametric phase diagrams in dS space \cite{balale}. The $U_o$ is from $10^{-2}$ to $2.50$ and $\Phi_o$ is from $0$ to $-14.00$. The solutions exist only on the presented curves unlike the case in AdS space. In other words, the points out of the curves do not satisfy the boundary conditions. We do not display the solutions with oscillation more than $i$ equal to $4$ for the $Z_2$-symmetric solution, denoted as $z_i$, because of the numerical difficulty. The $Z_2$-parity of $z_i$ is $(-1)^i$. We illustrate $z_1$ solutions with the red line, $z_2$ with the green, $z_3$ with the blue, $z_4$ with the green, $a_1$ with the black, and $a_3$ with the gray. The number of oscillations is decreased as $U_o$ or $\kappa$ is increased. For example, if we fix $U_o=1.50$ and $\kappa=0.30$ then only the $z_1$ solution is possible. We note that there exist closed loops composed of $a_{iL}$ and $a_{iR}$ solutions. For example, $a_{1L}$ and $a_{1R}$ solutions make a closed loop, which meets with the $z_1$ solution at a certain $U_o$ and $\Phi_o$. This point is known as a bifurcation point for $Z_2$-asymmetric solutions \cite{hawe, balale}. The curves move to the left and downward as $\kappa$ is increased. The size of the closed loop composed of $a_{1L}$ and $a_{1R}$ solutions is decreased as $\kappa$ is increased. We can see the second closed loop and the bifurcation point for the $a_3$ solution, i.e.\ the branch which $i>1$ in Fig.\ \ref{Fig.dS_phase}(a). We expect that such a closed loop and bifurcation point can be observed for the higher $i$ branches as the value of $\kappa$ is decreased.

From these parametric phase diagrams, we can estimate a parametric phase diagram or type of solution in the flat Minkowski space by limiting $\kappa$ to be zero. Let us consider taking $\kappa$ to be zero from the solutions in AdS space. In this case, the $b_1$ solution goes to the right and downward direction as the value of $\kappa$ is decreased. Then, the $s_1$ solution space is expanded by this effect. Finally we can expect that whole phase space will be covered by only the $s_1$ solution at $\kappa$ equal to zero. This expectation gives the exactly same result which is obtained by Fubini in the absence of gravity \cite{fubi01}.

\section{Summary and Discussion \label{Sec.5}}

There are three different kinds of bounce solutions. They are vacuum bubbles, oscillating bounce solutions, and the HM instanton, respectively. Oscillating bounce solutions and HM instanton are possible only if gravity is taken into account. In this paper, we have investigated oscillating Fubini bounce solutions of a self-gravitating scalar field under a tachyonic quartic potential and constructed the parametric phase diagrams. This is the extension of our previous paper \cite{bwdd}. The Fubini instanton describes the decay of a state located at the top of the potential to an arbitrary state through the tunneling. This solution connects one point at the center of the solution to a lower energy state through a thick wall.

The Euclidean bounce trajectory $\Phi_b(\eta)$ is constrained by the specific boundary conditions at $\eta=0$ and $\eta=\eta_{max}$. The evolution parameter $\eta_{max}$ is finite for dS space, while it is infinite for flat and AdS spaces. In AdS space, the late-time behavior of marginal solutions, $b_j$ solutions, is totally different from those of other solutions, $s_j$ solutions. The marginal solution reaches at $\Phi$ equal to $0$ essentially within a finite time, while other solutions reach at $\Phi$ equal to $0$ when $\eta$ goes to infinity. These late-time behaviors affect the tunneling probabilities. The action difference $B$ is finite for the marginal solution $b_j$, while the action difference $B$ diverges for other solutions $s_j$. In other words, the tunneling probabilities for the marginal solutions are finite, while the tunneling probabilities for other solutions are suppressed. In dS space, there exist $Z_2$-symmetric solutions \cite{bwdd, balale}. This is because the geometry of Euclidean dS space is invariant under the transformation $\eta \rightarrow \eta_{max} - \eta$. We have shown that there exist $Z_2$-asymmetric $a_3$ solutions. We expect that the closed loop composed of $a_{iL}$ and $a_{iR}$ solutions and bifurcation point can be observed for the higher $i$ branches as the value of $\kappa$ is decreased. There exist solutions stopped near the point zero ($\Phi=0$). In other words, the tunneling occurs from the point near the top of the potential to a certain state in dS space. It might be described by using thermal interpretation \cite{brwein}. Another interpretation is that dS spacetime has a gravitationally repulsive property which may prevent a full stop at $\Phi$ equal to $0$.

We have presented the parametric phase diagrams of the oscillating solutions with respect to three parameters ($\kappa$, $U_o$, and $\Phi_o$) in AdS and dS spaces. Of particular significance is that there always exist solutions in all parameter space in AdS space. The regions are divided depending on the number of oscillations. The solutions fill up the volume composed of three parameters. On the other hand, dS space allows solutions with codimension-one in parameter space. In other words, the solutions fill up the area composed of three parameters. Therefore, the solutions are more rich in AdS space than those in dS space. The dS space has compact geometry and the corresponding boundary conditions, which restrict the solution space. AdS spacetime has a gravitationally attractive property which may allow the wide range of a solution space, i.e.\ it promotes a full stop at $\Phi=0$.

From these parametric phase diagrams, we can estimate a parametric phase diagram or type of solutions in the flat Minkowski space by limiting $\kappa$ to be zero. Let us consider taking $\kappa$ to be zero from the solutions in AdS space. In this case, the $b_1$ solution goes to the right and downward direction as the value of $\kappa$ is decreased. Then, the $s_1$ solution space is expanded by this effect. Finally we can expect that whole phase space will be covered by only the $s_1$ solution at $\kappa$ equal to zero. This expectation gives exactly the same result which is obtained by Fubini in the absence of gravity \cite{fubi01}.

The issue on the instability of a tachyonic vacuum is very subtle. There are three possible modes, i.e. quantum fluctuation, the typical decay mode representing the homogeneous rolling on the potential, and the tunneling without a barrier \cite{kmwe, kml} as inhomogeneous transition. We did not consider quantum fluctuation in the present paper. The possibility may depend on both the curvature of the tachyonic top and the cosmological constant. If the shape of the potential near its top is very sharp we could not expect the existence of the tunneling without a barrier. If the shape near its top is flat enough, there exist two decay modes for the rolling and the tunneling. If so there is a competition between them. As pointed out in \cite{kml}, if the size of a solution is much smaller than the horizon size, the decay mode by the tunneling is distinguishable from others, while if the sizes are comparable, the decay mode by the tunneling becomes indistinguishable from others. To analyze which mode is dominant, one may need to estimate the decay time for each mode. If the decay time for the rolling is less than that for the tunneling the rolling becomes the dominant decay mode, and vice versa. In general all modes mix together. All modes may affect the cosmological large scale structure formation and CMB. To reach more understanding of the issue with all modes, further studies are needed. We will leave this issue as a future paper. In AdS space, a tachyonic top is known as perturbatively stable one without causing an instability of the background if the mass squared is at or above the Breitenlohner and Freedman bound \cite{brfr, brfr0}. Then the tunneling without a barrier is worthwhile to be explored as the only decay mode of a tachyonic vacuum.

\section*{Acknowledgements}

We would like to thank Remo Ruffini, Sung-Won Kim, Hyung Won Lee, Myeong-Gu Park, and She-Sheng Xue for their hospitality at the $13th$ Italian-Korean Symposium on Relativistic Astrophysics in Seoul, Korea, 15-19 July 2013. We would like to thank Misao Sasaki for his hospitality during our visit to Yukawa Institute for Theoretical Physics, Kyoto University. We would like to thank Kyoung Yee Kim and Kyung Kiu Kim for helpful discussions and comments. This work was supported by the National Research Foundation of Korea(NRF) grant funded by the Korea government(MSIP) (No. 2014R1A2A1A01002306). W.L. was supported by Basic Science Research Program through the National Research Foundation of Korea(NRF) funded by the Ministry of Education, Science and Technology(2012R1A1A2043908). D.Y. was supported by the JSPS Grant-in-Aid for Scientific Research (A) No. 21244033. D.Y. also would like to thank Leung Center for Cosmology and Particle Astrophysics (LeCosPA) of National Taiwan University (103R4000).

\newpage

\appendix
\section*{Appendix}{\bf Numerical proof for the finiteness of the exponent $B$}\label{sec:appa}

\begin{figure}[h]
  \centering
  \includegraphics[width=0.9\textwidth]{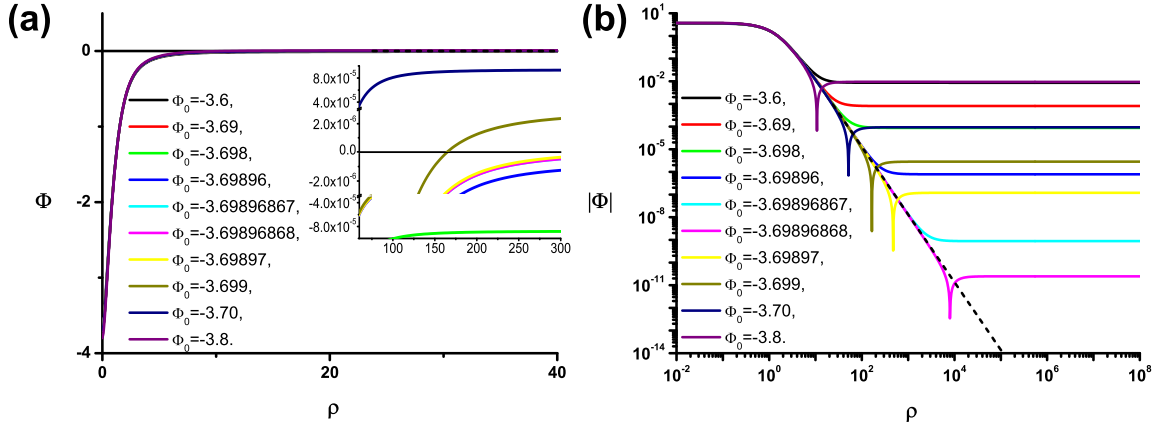}
  \caption{(color online). (a) The plot of $\Phi$ versus $\rho$ for five $s_1$ solutions and five $s_2$ solutions with $\Phi_o$ from $=-3.6$ to $=-3.8$ and (b) log-log scale plot of $|\Phi|$. The dashed line indicates the marginal solution. We choose $\kappa=0.30$ and $U_o=-0.30$.}
  \label{check-more}
\end{figure}

In Sec.\ \ref{Sec.3.2}, we have presented the finiteness of the exponent $B$ for the marginal solutions. In order to show the finiteness of the exponent $B$, we perform the numerical computation more carefully. We take $\kappa=0.30$ and $U_o=-0.30$.

First, we examine the behavior of the solution $\Phi$ as a function of $\rho$ in the log-log scale plot in more detail. We search for the numerical solutions in the small range of the initial values $\Phi_o$ between $-3.6$ to $-3.8$ around the marginal solution, $b_1$. Figure \ref{check-more} shows the plot of $\Phi$ as a function of $\rho$. Figure \ref{check-more}(a) presents the plot of $\Phi$, in which the right box in the same figure shows the magnification of a small region representing the late behavior of ten solutions. Only seven lines are shown in this figure. We insert two breaks, $3\times10^{-5} \sim 3\times10^{-6}$ and $-3\times10^{-5} \sim -3\times10^{-6}$. The broken line corresponds to overlapping five lines. After a break, $-3\times10^{-5} \sim -3\times10^{-6}$, the line is divided into three lines. The middle line corresponds to overlapping three lines with $\Phi_o$ among $-3.69896867$ and $-3.69897$. The line of $\Phi_o$ equal to $-3.8$ crosses over $\Phi$ equal to $0$ around $\rho \simeq 10$, the line of $\Phi_o$ equal to $-3.70$ crosses over $\Phi$ equal to $0$ around $\rho\simeq 50$, and the line of $\Phi_o$ equal to $-3.699$ crosses over $\Phi$ equal to $0$ around $\rho\simeq 170$, while the remaining seven lines are still within the negative region among ten lines in the graph. Figure \ref{check-more}(b) presents the log-log scale plot of $|\Phi|$ with $\rho$. The solutions are divided into two groups depending on whether they have the downward peak or not. Since we plot the absolute value of $\Phi$, the downward peaks indicate the region where the sign of a field changes. They represent the behavior coming back to $\Phi$ equal to $0$ after oscillation in numerical computation. The upper cases in the plot legend correspond to the $s_1$ solution, while the lower five cases to the $s_2$ solutions. The dashed line indicates the marginal solution, which has the initial values $\Phi_o$ between $-3.69896867$ and $-3.69896868$. In this figure, all the solutions show a certain behavior. The fields are slowly decreasing at the first stage, and after that they are linearly decreasing. The later behavior of solutions is quite different. The splitting occurs along the dashed line as it approaches the initial value of $\Phi_o$ for the marginal solution. That is, the splitting point moves downward as the initial value of $|\Phi_o|$ is increased among $s_1$ solutions, while the splitting point moves upward as the initial value is increased among $s_2$ solutions after over the initial value for the marginal solution. In other words, the field with $\Phi_o$ equal to $-3.6$ rolls down the inverted potential quickly and is slowly approaching to $\Phi$ equal to $0$ after splitting from the dashed line. If the initial value is increased to $\Phi_o$ equal to $-3.69$, the field trajectory is approaching a little more and slows down a little later. The splitting occurs a little later. If the initial value is over that for the marginal solution, the splitting occurs in the opposite way to the above cases. This kind of behavior represents the existence of the marginal solution as the undershoot-overshoot numerical procedure.

\begin{figure}
  \centering
  \includegraphics[width=0.9\textwidth]{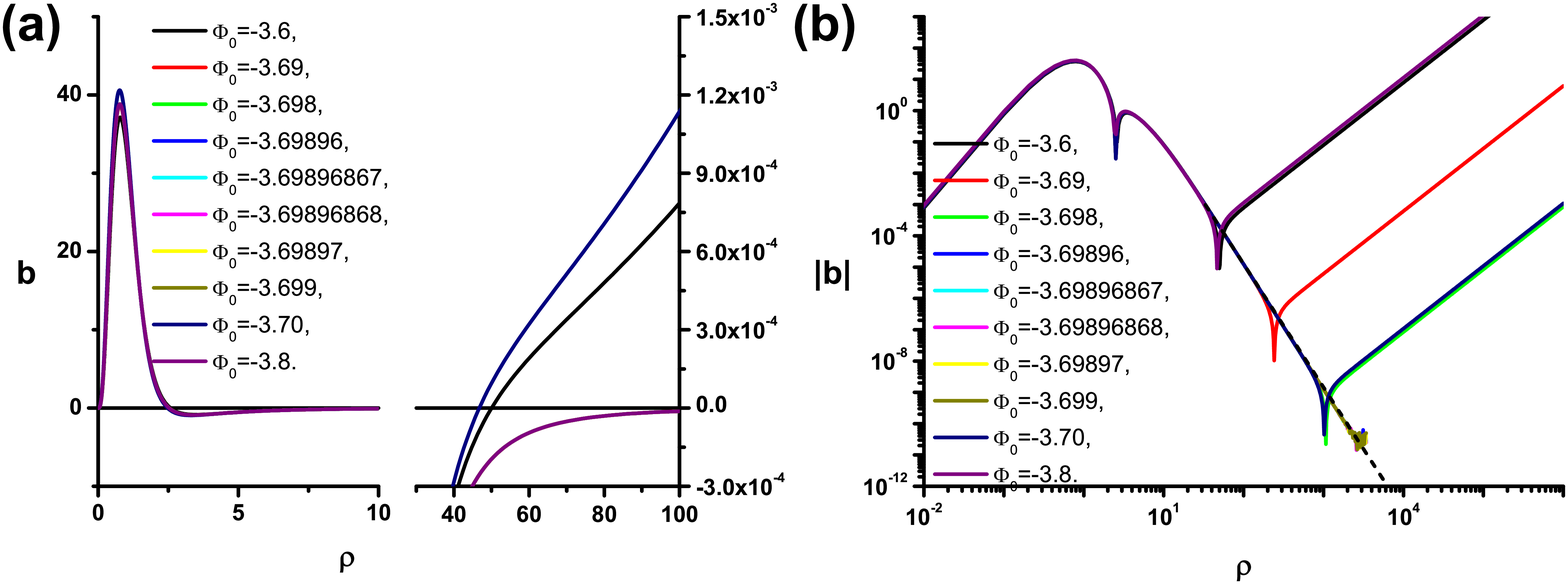}
  \caption{(color online). (a) The plot of the integrand $b$  and (b) Log-log scale plot of $|b|$. }
  \label{check-AdS-action}
\end{figure}

Second, we numerically evaluate the exponent $B$. Figure \ref{check-AdS-action} shows the plot of the integrand $b$ as a function of $\rho$ for ten solutions. Figure \ref{check-AdS-action}(a) presents the plot of $b$. The upper five lines in the plot legend correspond to the $s_1$ solutions, while the lower five lines to the $s_2$ solutions. The $b$s have the maximum value at the first stage, cross over to the negative region, and cross over again to the positive region except for the marginal solution. For the marginal solution, the value of $b$ is asymptotically approaching to zero. The right figure in the same figure (a) shows the magnification of a small region representing the late behavior of the solutions. The two lines crossing over to the positive region are shown in the figure. Figure \ref{check-AdS-action}(b) presents the log-log scale plot of the integrand $|b|$ as a function of $\rho$ for ten solutions. From Eq.\ (\ref{semiexpob}), the integrand $b$ is given by
\begin{equation}
2\pi^2 \left[\frac{\rho^3 [-U(bs)]}{\sqrt{1+\frac{\kappa\rho^2}{3}[\frac{1}{2}\Phi'^2-U(bs)] }} -  \frac{\rho^3 [-U(bg)]}{\sqrt{1+\frac{\kappa\rho^2}{3}[-U(bg)] }} \right]\,.  \label{integrandb}
\end{equation}
The first downward peak represents the value of $b$ going to the negative value. The second downward peak represents the value of $b$ turning back to the positive value. The five cases with initial values $\Phi_o$ between $-3.69896$ and $-3.699$ are overlapped. These overlapping lines are approaching the zero value. In Fig.\ \ref{check-AdS-action}(b), the overlapping lines are linearly decreasing with the slope about $-4$. Others are separated from the overlapping lines after the second downward peak. They have the slope about $+2$ as the late behavior. Thus, the integrand for those solutions seems to diverge. The integrand $b$ of the marginal solution has the slope $-4$, we can integrate directly within the range $(\rho_1, \rho_2)$ as
\begin{equation}
B\simeq \int^{\rho_2}_{\rho_1} \rho^{-4} d\rho = \frac{1}{3}(\rho^{-3}_1 -\rho^{-3}_2)\,,
\end{equation}
where the $\rho_1$ and $\rho_2$ terms indicate the integration of each square bracket in Eq.\ (\ref{integrandb}). We take $\rho_2=10^{11}$, then the $\rho_2$ term has an extremely small value. Thus, we do not need to consider the integration range for the maximum $\rho$ because it gives a negligible effect.

In this point, it is difficult to divide the overlapping lines into each line due to the limitation of the precision in the numerical calculation. We include the wad of ending region in overlapping lines. There seems to be an error after the wad. We leave the graph without splitting for the overlapped solutions.

\begin{figure}
  \centering
  \includegraphics[width=0.5\textwidth]{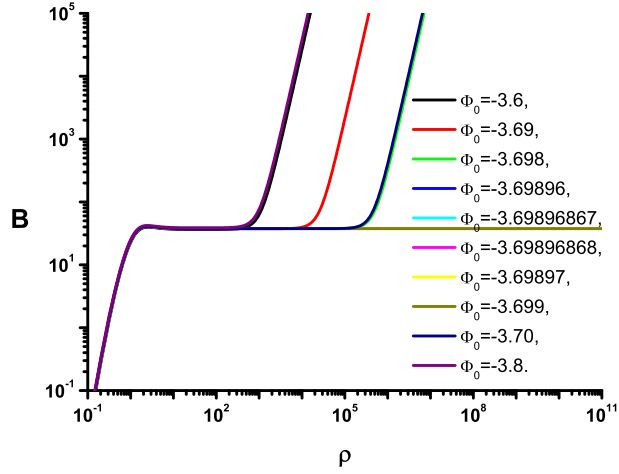}
  \caption{(color online). The exponent $B$ as a function of $\rho$. }
  \label{Action_difference_B}
\end{figure}

Figure \ref{Action_difference_B} shows the exponent $B$ as a function of $\rho$. The graph shows the existence of the constant line, which indicates the finiteness of $B$ for the marginal solution. The splitting occurs along the constant line as it approaches the initial value of $\Phi$ for the marginal solution. The five lines with initial values $\Phi_o$ between $-3.69896$ and $-3.699$ are overlapped in the constant line. We expect that the exponent $B$ for these overlapping solutions will diverge at very large $\rho$ except for the exact marginal solution. This kind of behavior for the marginal solution represents the existence of the finite $B$ to give the finite probability.

\end{document}